# Ultrafast Demagnetization Governed by Spin Fluctuations in $CaRuO_3/SrTiO_3$ Superlattice

Yu-Han Gao[1,2], Wen-Xiao Shi[1,2], Yuan-Sha Chen[1,2], Ji-Rong Sun[1,2], Qing-Lin Yang[1,2], Xu Yang[1,3], Zhuo Deng[1,2], Peng-Tao Yang[1,2], Zheng Chang[1,2], Hong-Mei Feng[3], Wei He[1], Xiang-Qun Zhang[1] and Zhao-Hua Cheng[1,2,3,*]

[1]Beijing National Laboratory for Condensed Matter Physics, Institute of Physics, Chinese Academy of Sciences, Beijing 100190, China;

[2]School of Physical Sciences, University of Chinese Academy of Sciences, Beijing 100049, China;

[3]Songshan Lake Materials Laboratory, Dongguan, Guangdong 523808, China.

[*]Corresponding authors: zhcheng@iphy.ac.cn

## Abstract

**For ultrafast magnetization switching devices, critical slowing down in conventional ferromagnets near their Curie temperature constitutes a key challenge that must be overcome. In contrast to this typical behavior, we observe an anomalous acceleration of demagnetization in $CaRuO_3/SrTiO_3$ superlattices, a moderately correlated weak itinerant ferromagnet. The demagnetization rate increases with rising temperature, pump fluence, and applied magnetic field. To explain these anomalous phenomena, we develop a phenomenological model integrating the three-temperature model with self-consistent renormalization theory. Because the intrinsic gradient magnetism of the superlattice suppresses the typical divergence of specific heat, the conventional thermodynamic bottleneck is bypassed. Our model reveals that this decoupling enables the ultrafast dynamics to be predominantly governed by the spin-fluctuation-driven enhancement of the electron-spin scattering vertex. Our work demonstrates how spatial inhomogeneity can decouple macroscopic thermodynamic singularities from microscopic scattering processes, offering a new paradigm for manipulating ultrafast spin dynamics in correlated quantum materials. The pronounced sensitivity of the demagnetization rate to external parameters further suggests the potential for designing highly tunable ultrafast spintronic devices that leverage**

**enhanced fluctuations near the magnetic instability.**

## Introduction

The ultrafast manipulation of magnetic order represents a central goal of modern spintronics and a vital frontier for investigating non-equilibrium dynamics in condensed matter physics [1-4]. While interest has intensified since the discovery of ultrafast demagnetization in nickel [5] and other systems [6-11], the fundamental microscopic mechanisms driving these processes remain fiercely debated [12-18]. While phenomenological frameworks like the three-temperature model (3TM) describe macroscopic energy transfer [5], identifying the specific microscopic spin-flip channels remains a primary objective [19-21]. Unraveling these mechanisms is critical for addressing intrinsic limitations in spintronics, such as the critical slowing down near phase transitions [22,23]. Beyond non-local transport [11,24], intrinsic demagnetization is conventionally attributed to two local channels. The first involves spin-orbit coupling mediated Elliott-Yafet scattering [25,26], such as the microscopic three-temperature model (M3TM) [13]. Although successful in describing macroscopic demagnetization via the mean-field collapse of the exchange gap, their spin-flip events rely on non-magnetic scattering [9,13,27]. The second emphasizes the rapid redistribution of energy and angular momentum by exchange interactions, including Stoner excitations [16,20] and electron-magnon scattering [12,28,29]. However, these conventional frameworks rarely treat the mode-mode coupling between spin fluctuations, which provides a crucial vertex renormalization channel for electrons. Such strongly coupled collective fluctuations dominate itinerant systems near magnetic instability [30,31], driving anomalous scattering rates that cannot be captured by conventional non-interacting magnon theories alone [32,33]. Their critical role is corroborated by recent phase transition experiments [34] and theoretical demonstrations of light-induced nonequilibrium spin fluctuations [35]. Thus, integrating these strongly coupled spin fluctuations into the ultrafast demagnetization paradigm is essential for explaining novel phenomena in correlated quantum materials and completing a comprehensive microscopic picture of ultrafast spin dynamics.

Recent studies have reported the emergence of a well-defined ferromagnetic (FM) ground state in heterostructures comprising two non-magnetic constituents with significant symmetry mismatch [36,37]. This interfacial ferromagnetism arises in $[(CaRuO_3)_n/(SrTiO_3)_m]_x$ superlattices (CRO/STO SLs), comprising $n$ unit-cell (uc) CRO and $m$ uc STO layers, repeated $x$ times. Unless otherwise noted, all experimental data presented in this study are from the $[CRO_{10}/STO_1]_{10}$ superlattice, which exhibits a Curie temperature $T_C = 90$ K and a saturation magnetization $M_s = 0.36\ \mu_B/\text{f.u.}$ [37]. The system behaves as a moderately correlated weak itinerant ferromagnet [38,39], with a magnetic ground state situated near the quantum critical point (QCP) [40]. The proximity to the QCP, inherent electronic correlations, and quasi-two-dimensional (quasi-2D) magnetic confinement collectively amplify intense spin fluctuations, aligning with the Mermin-Wagner theorem [41,42]. Furthermore, unlike uniform ferromagnets, the magnetic order in CRO/STO SLs induced by interfacial symmetry breaking, exhibits a spatial gradient that decays away from the interfaces [36,37], as shown in the right panel of Fig.1 (a). This depth-dependent distribution of the order parameter effectively smears the macroscopic phase transition and suppresses the characteristic divergence of specific heat near the $T_C$ [43].

Here, we systematically investigate ultrafast spin dynamics in CRO/STO SLs via time-resolved pump-probe measurements dependent on temperature, pump fluence and magnetic field. While the critical slowing down near $T_C$ typically imposes a fundamental speed limit on magnetic switching devices [44], our experiments reveal a significant departure from this constraint. As the temperature rises, the demagnetization process accelerates continuously, effectively bypassing this thermodynamic bottleneck. Unlike conventional ferromagnets, the demagnetization rate here increases pronouncedly with increasing pump fluence and magnetic field, demonstrating sensitive tunability. To explain these results, we develop a phenomenological model integrating self-consistent renormalization (SCR) theory. The model successfully captures the observed anomalies, revealing that the system's inherent gradient magnetism unmasks the spin-fluctuation-driven scattering enhancement, allowing it to dominate the dynamics. Our work provides new insight into non-equilibrium spin dynamics in correlated materials with strong coupled fluctuations and demonstrates a pathway to actively manipulate ultrafast demagnetization using multiple external parameters.

## Results and Discussion

### A. Ultrafast Demagnetization Dynamics in SLs

To investigate the ultrafast spin dynamics of the CRO/STO SLs, time-resolved magneto-optical Kerr effect (TR-MOKE) and transient reflectivity measurements were performed using an ultrafast all-optical pump-probe system. The schematic of the experimental geometry is shown in the left panel of Fig. 1(a). Fig. 1(b) displays representative ultrafast demagnetization curves under varying temperature ($T$), pump fluence ($F$), and magnetic field ($H$), with non-magnetic artifacts removed via field-reversal symmetry subtraction (see Supplemental Material Sec. S1). These dynamics are strongly modulated by $T$, $F$, and $H$, evolving most significantly within the initial $\sim$3 ps. Specifically, conditions of low-$T$ and low-$F$ (black symbols: 20 K, 0.35 mJ·cm$^{-2}$, 5.8 kOe) yield a slow, two-step (Type-II) demagnetization process. Conversely, high-$T$ and high-$F$ conditions (red symbols: 90 K, 1.38 mJ·cm$^{-2}$, 18.5 kOe) trigger a distinctly faster, one-step (Type-I) demagnetization process. An intermediate regime (blue symbols: 20 K, 1.38 mJ·cm$^{-2}$, 18.5 kOe) continuously bridges these two distinct kinetic behaviors. While sensitivity to external parameters is a general feature of ultrafast magnetism, the specific trends observed here deviate from the conventional Type-I to Type-II transition [9], manifesting an anomalous crossover from slow Type-II to fast Type-I dynamics with increasing excitation. This distinct response implies an unconventional microscopic mechanism governing the ultrafast demagnetization in this system.

To quantitatively analyze the demagnetization dynamics, the TR-MOKE data are fitted using the 3TM [5]. This model treats the ferromagnet as three coupled subsystems: electrons, phonons, and spins, each of which has its own effective temperature ($T_\mathrm{e}$, $T_\mathrm{p}$and $T_\mathrm{s}$) [45]. Assuming that the transient reflectivity change $|\Delta R/R| \propto \Delta T_\mathrm{e}$ and the MO signal change $|\Delta\mathrm{MO}| \propto (-\Delta M_z/M_z) \propto \Delta T_\mathrm{s}$ [5], we fitted the MO signals [solid lines in Fig. 1(b)] using the following relation:

$$\frac{\mathrm{d}T_\mathrm{s}}{\mathrm{d}t} = \Gamma_\mathrm{s}[T_\mathrm{e}(t) - T_\mathrm{s}], \tag{1}$$

where $T_\mathrm{e}(t)$ is approximated by fitting the reflectivity curves, measured under identical conditions as the MO signals [see Figs. S2(a), S3(a), S4], with a bi-exponential function [46]: $\Delta R/R = A_1 \exp(-t/\tau_1) + A_2 \exp(-t/\tau_2)$. Both reflectivity and MO signal fits are convoluted with a Lorentzian laser pulse profile. To account for thermal diffusion, a prefactor $\exp(-t/\tau_\mathrm{D})$ is applied to $T_\mathrm{s}(t)$. The contribution of spin–phonon interaction term $G_\mathrm{sp}(T_\mathrm{p} - T_\mathrm{s})$ is considered negligible compared to the electron-spin (e-s) interaction on the timescale $< 10$ ps [9,15]. We define the spin relaxation rate as $\Gamma_\mathrm{s} \equiv G_\mathrm{es}/C_\mathrm{s}$, where $G_\mathrm{es}$ denotes the e-s interaction and $C_\mathrm{s}$ is the spin specific heat. Within this framework, $\Gamma_\mathrm{s}$ is the sole parameter governing the evolution of $T_\mathrm{s}$ according to Eq. (1), thereby providing a direct quantitative measure of the demagnetization speed.

To further clarify the variations in demagnetization characteristics and to explore the underlying microscopic mechanism, we systematically measured TR-MOKE dynamics across varying *T*, *F*, and *H*. As shown in Fig. 2(a), under fixed *H* and *F*, the demagnetization dynamics exhibit a clear temperature dependence. The corresponding reflectivity data used for the fitting are presented in Fig. S2. In Fig. 2(a), the time delay ($t_\mathrm{max}$) at which the MO signal change reaches its maximum is marked by star symbols. To quantify the demagnetization speed, we extracted the characteristic demagnetization time $\tau_\mathrm{M}$, using a phenomenological fitting function (see Supplemental Material Sec. S5 for details). The acceleration of demagnetization is visualized in Fig. 2(d), which plots the temperature dependence of both $t_\mathrm{max}$ and $\tau_\mathrm{M}$. For $T < T_C(\sim 90\ \mathrm{K})$, $t_\mathrm{max}$ and $\tau_\mathrm{M}$ decrease monotonically with increasing temperature, indicating a faster demagnetization process. Above $T_\mathrm{C}$, both $t_\mathrm{max}$ and $\tau_\mathrm{M}$ remain nearly constant, suggesting that the acceleration of demagnetization saturates. The persistence of the transient MO signal well above $T_\mathrm{C}$ is attributed to the *H*-induced paramagnetic response of localized magnetic clusters originating from interfacial Sr/Ca intermixing [36]. Notably, the spin relaxation rate saturates above $T_\mathrm{C}$, as detailed and experimentally verified in Supplemental Material Sec. S12. Similarly, Fig. 2(b) and 2(c) demonstrate that demagnetization accelerates with increasing $F$ and $H$ ($T$ = 20 K), respectively. The corresponding decreasing trends of $t_{\mathrm{max}}$ and $\tau_{\mathrm{M}}$ are plotted in Fig. 2(e, f). For the $F$-dependent (or $H$-dependent) measurements at 20 K, two distinct fixed values of $H$ (or $F$) are selected to confirm the robustness of the results [see Fig. S6]. These observations were further validated by measurements on a $[CRO_{10}/STO_3]_{10}$ SL [see Fig. S7]. The spin relaxation rates $\Gamma_{\mathrm{s}}$, extracted from fitting the MO signals, will be presented in the following section in conjunction with the theoretical analysis.

Significantly, the spin dynamics observed in CRO/STO SLs are highly anomalous. In most $T$-dependent ultrafast demagnetization studies, the demagnetization process slows down as the temperature approaches $T_{\mathrm{C}}$ [22,47], the phenomenon known as critical slowing down, often accompanied by a crossover from Type-I to Type-II characteristics [9,48]. The increasing $F$ is often considered to be equivalent to elevating $T$ experimentally, resulting in similar trends [9,13]. In contrast, our results exhibit the opposite trend: an acceleration of demagnetization as the temperature rises, persisting over a wide temperature range, concurrent with a transition from Type-II to Type-I behavior. Furthermore, we find that the spin relaxation rate in this system can be sensitively manipulated, showing a pronounced increase while increasing $F$ or $H$. This counter-intuitive trend necessitates a deeper theoretical analysis that explicitly incorporates the unique properties of the CRO/STO SLs.

## B. Microscopic Mechanism of Spin-Fluctuation-Governed Demagnetization

To describe the ultrafast dynamics of such system, the non-negligible impact of coupled spin fluctuations must be explicitly addressed. We therefore start with the single-band Hubbard model to capture the core physics of itinerant ferromagnetism and electronic correlations [49]. By applying self-consistent renormalization (SCR) theory [50], we treat the Coulomb interaction beyond mean-field approximation to incorporate mode-mode coupling of spin fluctuations. This mode-mode coupling, intrinsic to the SCR approach, identifies spin fluctuations as the dominant scattering channel responsible for the ultrafast demagnetization dynamics in the strong fluctuation regime.

Within the 3TM framework, the spin relaxation rate is defined as $\Gamma_{\mathrm{s}} = G_{\mathrm{es}}/C_{\mathrm{s}}$. In itinerant systems near magnetic instability, longitudinal modes become notably soft and strongly couple with transverse modes [30,31]. We therefore focus on the scattering channel heavily renormalized by these spin fluctuations, denoting the resulting relaxation rate as $\Gamma_{\mathrm{s}}^{\mathrm{fluc}}$. By deriving the energy transfer rate between itinerant electrons and the spin fluctuation system (see Supplemental Material Sec. S14 for details), we obtain the expression for the electron-spin fluctuation (e-sf) interaction coefficient as

$$G_{\mathrm{es}}^{\mathrm{fluc}} = 2|M'|^2 D_{\uparrow} D_{\downarrow} \sum_{\mathbf{q}} \int_0^{\infty} \mathrm{d}\omega\, (\omega^2) \mathrm{Im}[\chi(\mathbf{q},\omega)] \frac{\partial n_{\mathrm{B}}(\omega, T_s)}{\partial T_s}, \tag{2}$$

where $|M'|^2$ is the effective e-sf scattering matrix element, $D_{\uparrow(\downarrow)}$ is the density of states at the Fermi surface for spin-up (down) electrons, $n_{\mathrm{B}}$ is the Bose-Einstein distribution, and $\mathrm{Im}[\chi(\mathbf{q},\omega)]$ represents the imaginary part of the magnetic susceptibility, where $\mathbf{q}$ and $\omega$ denote the wave vector and frequency of spin fluctuations, respectively.

In conventional Stoner or Heisenberg models, the scattering matrix element is typically derived from the Coulomb interaction in the random phase approximation, treated as a $T$-independent constant: $|M_0|^2 \propto U^2 = \mathrm{const}$, where $U$ denotes the Coulomb interaction [50]. However, this approximation breaks down in regimes dominated by strong fluctuations. As described by the SCR theory, distinct spin fluctuation modes are not independent; they interact and scatter with one another through mode-mode coupling [51]. This mode-mode coupling renormalizes not only the spin fluctuation propagator but also the interaction vertex between itinerant electrons and spin fluctuations. We account for this by assuming that the effective scattering matrix element $|M'|^2$ is renormalized by thermal spin fluctuations: $|M'|^2 \approx \lambda\langle S^2\rangle_T^2$. Here, $\lambda$ is a coupling constant, and $\langle S^2\rangle_T$ is the mean-square amplitude of thermal spin fluctuations. As the temperature increases, $\langle S^2\rangle_T$ increases significantly, leading to a thermal enhancement of the e-sf scattering strength. By combining the results of the quasi-2D SCR theory and the integration of the imaginary part of the susceptibility over the spin fluctuation spectrum [52], the e-sf interaction can be expressed as

$$G_{\mathrm{es}}^{\mathrm{fluc}}(y,t) \propto \frac{(4-\varepsilon^3)}{(3-\varepsilon^2)^3}[G(y,\varepsilon)]^2 t^3,$$
$$G(y,\varepsilon) = y\left(1+\frac{1}{\varepsilon^2}\right) - \frac{\pi}{\varepsilon}\sqrt{y} + 2 - 2\ln\varepsilon\,. \tag{3}$$

Under high-$T$ and long-range correlation approximations, the dynamic scaling of $G_{\mathrm{es}}^{\mathrm{fluc}}$ is uniquely determined by the reduced temperature $t = T/T_0$ ( $T_0$ being the characteristic frequency scale), the dimensionless inverse susceptibility $y$, and the structural dimensionality factor $\varepsilon$ (see Supplemental Material Sec. S14 for details) [52]. For a given material, $\varepsilon$ is constant, implying that $t$ and $y$ determine the behavior of $G_{\mathrm{es}}^{\mathrm{fluc}}$. In conventional ferromagnets, the renormalized vertex correction is appreciable only in the strong-coupling and critical regime, which corresponds to a very narrow temperature range near $T_{\mathrm{C}}$. In contrast, our SLs exhibit intrinsically strong fluctuations and a pronounced gradient magnetism below $T_{\mathrm{C}}$, which considerably expand the temperature interval over which the renormalization of $|M'|^2$ remains effective. Consequently, the expression above can serve as an effective description for the ordered and crossover regimes where itinerant spin fluctuations dominate.

The spin relaxation rate $\Gamma_{\mathrm{s}}^{\mathrm{fluc}}$ depends not only on $G_{\mathrm{es}}^{\mathrm{fluc}}$ but also on the specific heat $C_{\mathrm{s}}^{\mathrm{fluc}}$, which appears in the denominator. By invoking the same approximations employed for $G_{\mathrm{es}}^{\mathrm{fluc}}$, we derive the relation for $C_{\mathrm{s}}^{\mathrm{fluc}}$ as follows

$$C_{\mathrm{s}}^{\mathrm{fluc}}(y,t) \propto \left(\frac{1}{3}-\frac{1}{12}\varepsilon^3\right)\ln t - Y(y,\varepsilon) + \mathrm{const},$$
$$Y(y,\varepsilon) = -y^3\frac{1}{4\varepsilon^3} + y^2\ln y\frac{1}{4\varepsilon} + y^2\left(\frac{2}{3}-\frac{2}{3}\frac{1}{\varepsilon}-\frac{\ln\varepsilon}{2\varepsilon}\right) + y\left(\frac{2}{3}-\frac{\varepsilon}{2}\right) + \frac{1}{3}\ln y\,. \tag{4}$$

The detailed description can be found in Supplemental Material Sec. S15. The $\ln y$ term of $Y(y,\varepsilon)$ in Eq. (4) is of particular importance. In conventional ferromagnets, the magnetic susceptibility diverges as the temperature approaches the Curie point $T_{\mathrm{C}}$ [53], causing the inverse susceptibility $y \to 0$ and $\ln y \to -\infty$. This $C_{\mathrm{s}}^{\mathrm{fluc}} \to \infty$ divergence forces $\Gamma_{\mathrm{s}}^{\mathrm{fluc}} \to 0$, creating the thermodynamic bottleneck responsible for critical slowing down. However, in our SLs, the gradient magnetism implies a spatial distribution of $M_{\mathrm{s}}$ and $T_{\mathrm{C}}$ [see in the right panel of Fig.1(a)]. This distribution prevents the macroscopic susceptibility from diverging, as shown in Fig. S8(b), keeping $y$ finite and thereby smearing out the divergence of $C_{\mathrm{s}}^{\mathrm{fluc}}$.

To intuitively illustrate the competition between the scattering and thermodynamic factors, we present a schematic model in Fig. 3. In conventional ferromagnets [Fig. 3(a)], the thermodynamic divergence of spin specific heat overwhelms the scattering term and suppresses the spin relaxation rate. This typifies critical slowing down, often observed experimentally as a decrease in demagnetization rate or a Type-I to Type-II transition [9,54]. In contrast, the CRO/STO SL [Fig. 3(b)] is not a homogeneous system, for which static magnetic measurements indicate a significant spatial distribution of $M_{\mathrm{s}}$ [36]. Similar to finite-size scaling where singularities are rounded off [55], the spatial gradient in the order parameter prevents the correlation length from diverging globally [56]. The gradient magnetism smears the specific heat divergence. Consequently, the intrinsic acceleration of spin-flip scattering driven by vertex renormalization is unmasked and dominates the dynamics, resulting in a continuous acceleration of $\Gamma_{\mathrm{s}}$ with rising temperature when $T < T_{\mathrm{C}}$. Far above $T_{\mathrm{C}}$, however, $\Gamma_{\mathrm{s}}$ plateaus at a constant asymptotic limit [see Fig. 3(b) and see Supplemental Material Sec. S12 for details]. This acceleration does not violate the universality of critical phenomena in continuous phase transitions.

Mathematically, combining the expressions for $G_{\mathrm{es}}^{\mathrm{fluc}}$ and $C_{\mathrm{s}}^{\mathrm{fluc}}$, the relation between $\Gamma_{\mathrm{s}}^{\mathrm{fluc}}$ and reduced parameters $(y, t)$ is derived as

$$\Gamma_{\mathrm{s}}^{\mathrm{fluc}}(y,t) \propto \frac{[G(y,\varepsilon)]^2 t^3}{\left(\frac{1}{3} - \frac{1}{12}\varepsilon^3\right)\ln t - Y(y,\varepsilon) + \mathrm{const}}. \quad (5)$$

The resulting expression captures the anomalous ultrafast demagnetization observed in our experiments. To quantitatively evaluate whether our spin-fluctuation framework can account for these behaviors, the $\Gamma_{\mathrm{s}}$ data are fitted using the $\Gamma_{\mathrm{s}}^{\mathrm{fluc}}(y,t)$ function in Eq. (5), which integrates the experimental $\chi^{-1} - T$ and $\chi^{-1} - H$ relations shown in Fig. S8(c, e). As shown by the colored lines in Fig. 4, the theoretical model exhibits excellent agreement with the experimental data, strictly reproducing the sharp rise of $\Gamma_{\mathrm{s}}$ with increasing $T$ [see in Fig. 4(a)]. The global fitting yields the dimensionality factor $\varepsilon \approx 0.498$, reflecting the quasi-2D magnetic confinement at the interfaces, and the spin-fluctuation characteristic temperature $T_0 \approx 398$ K, which falls within the expected energy scale for typical weak itinerant ferromagnets [57]. These physically compatible parameters suggest that incorporating spin-fluctuation scattering beyond the mean-field approximation is a viable approach to capture the observed dynamics, supporting the decoupling of thermodynamic singularity and scattering enhancement as a key microscopic mechanism of the observed anomalies. Furthermore, the observed $F$-dependence offers additional evidence for this fluctuation-driven thermal excitation mechanism. In the ultrafast regime, the pump pulse determines the peak transient electron temperature, creating a highly non-equilibrium state. Physically, this transient high-temperature environment is analogous to an elevated ambient temperature [13], which amplifies the spin fluctuation amplitude and further opens efficient angular momentum dissipation channels. This accounts for the systematic increase in $\Gamma_s$ with $F$.

The $H$-dependence in Fig. 4(c) allows us to explicitly distinguish our mechanism from conventional magnon-driven scattering. In a conventional magnon-driven scenario, an applied field $H$ induces a Zeeman gap in the magnon spectrum [15]. This gap effectively suppresses the thermal excitation of low-energy magnons, limiting the available scattering channels and consequently reducing the spin-flip probability [51]. In contrast, our experiments show that $\Gamma_s$ increases with $H$ [Fig. 4(c)], inconsistent with the simple magnon scenario. Furthermore, this significant $H$-dependence cannot be explained by conventional mean-field approaches such as M3TM, where the Zeeman splitting under our experimental fields yields an almost negligible modification to the spin-flip rate. This anomalous $H$-induced acceleration is naturally explained by the SCR theory for weak itinerant ferromagnets, where the mathematical form of the susceptibility used in our model [Eq. (S13)] describes the overdamped, diffusive spin fluctuations characteristic of such systems. In this framework, $H$ modulates the coupled spin fluctuation spectrum by increasing the dimensionless reciprocal magnetic susceptibility, $y$. Unlike the magnon suppression effect, the scattering rate $\Gamma_s^{\mathrm{fluc}}$ evolves nonlinearly with $y$ within the quasi-2D SCR model. We note that this trend refers to the experimentally accessible moderate-$H$ regime; in the high-$H$ limit, the overall suppression of spin fluctuations is expected to weaken the vertex-renormalization effect, as discussed in Supplemental Material Sec. S13. The agreement between this model and the observed acceleration supports that the ultrafast dynamics are governed by the spin-fluctuation-driven renormalization of the electron-spin scattering vertex.

## Conclusion

In summary, we observed a counterintuitive acceleration of ultrafast demagnetization in $CaRuO_3/SrTiO_3$ superlattices with increasing temperature, pump fluence, and magnetic field, defying the conventional behavior of critical slowing down in conventional ferromagnets. By integrating self-consistent renormalization theory with the three-temperature model framework, we demonstrate that this acceleration is governed by strongly enhanced electron-spin scattering driven by thermal spin fluctuations. This spin-fluctuation-governed mechanism is uniquely enabled by the system's gradient magnetism, revealing a general principle: in spatially inhomogeneous quantum materials, macroscopic thermodynamic singularities can be effectively decoupled from the microscopic scattering rate. By bypassing the traditional thermodynamic bottleneck, this insight provides a new pathway for manipulating complex correlated quantum matter. The high sensitivity of the demagnetization rate to external parameters such as temperature, optical excitation, and magnetic field, demonstrates the potential for designing highly tunable ultrafast spintronic devices that turn the enhanced fluctuations from a liability into an asset for boosting operational speeds.

## Methods

### Sample fabrications.

High-quality $[CRO_{10}/STO_n]_{10}$ superlattices (SLs) were synthesized via pulsed laser deposition. In these heterostructures, ferromagnetism is induced through interfacial symmetry mismatch. All samples were epitaxially grown on (111)-oriented $SrTiO_3$ substrates. Detailed structural characterization and static magnetic properties, which confirm the successful stabilization of a magnetic ground state, are reported in prior work [36,37]. We employ these SLs as a model system to investigate the role of strong spin fluctuations in ultrafast dynamics.

### TR-MOKE and Transient Reflectivity Measurements.

Time-resolved magneto-optical Kerr effect (TR-MOKE) and transient reflectivity measurements were performed on CRO/STO SLs using an ultrafast pump-probe system. The pump laser pulses were generated from a Ti:sapphire oscillator, with a center wavelength of 780 nm, a beam size of 14 μm, a repetition rate of 5.2 MHz and a pulse duration of about 110 fs. The 390 nm probe pulses were generated via second-harmonic generation of the 780 nm pulses with a $\beta$-barium borate (BBO) crystal. Both beams were incident normally onto the sample surface, with a magnetic field of up to 20 kOe applied along the z-axis. The sample was mounted in a liquid-helium-cooled cryostat with temperature stability within $\pm 0.01\ \mathrm{K \cdot h^{-1}}$. The pump laser pulse, which has a photon energy of 1.59 eV, is absorbed only by the metallic CRO layers, while the STO layers (band gap ~ 3.19 eV) are optically transparent at this photon energy. The estimated optical penetration depths in CRO are about 65 nm for the pump and 45 nm for the probe, both exceeding the total thickness of the SLs (~25 nm). Therefore, the pump pulses excite the entire SL, and the probe pulses also capture the magneto-optical response from all constituent CRO layers.

**Acknowledgments**

This work was supported by the National Key Research Program of China (grant No. 2022YFA1403302 and 2024YFA1408702) the National Natural Sciences Foundation of China (grant nos. U22A20115, and 52031015), and the Key Research Program of Frontier Sciences, CAS (Grant No. ZDYZ2012-2).

**Conflict of Interest**

The authors declare no conflict of interest.

**Author Contributions**

Z. H. C. conceived and supervised the project. Y. H. G. performed time-resolved pump-probe measurements and theoretical calculations. W. X. S., Y. S. C., and J. R. S. synthesized the samples. Q. L. Y., X. Y., Z. D., P. T. Y., Z. C. assisted with the pump-probe measurements. H. M. F. made significant contributions to static magnetic measurements. X. Q. Z. and W. H. contributed to the development and maintenance of the ultrafast optical system. Y. H. G. and Z. H. C. drafted the manuscript. All authors discussed the results and commented on the manuscript.

**Figure 1**

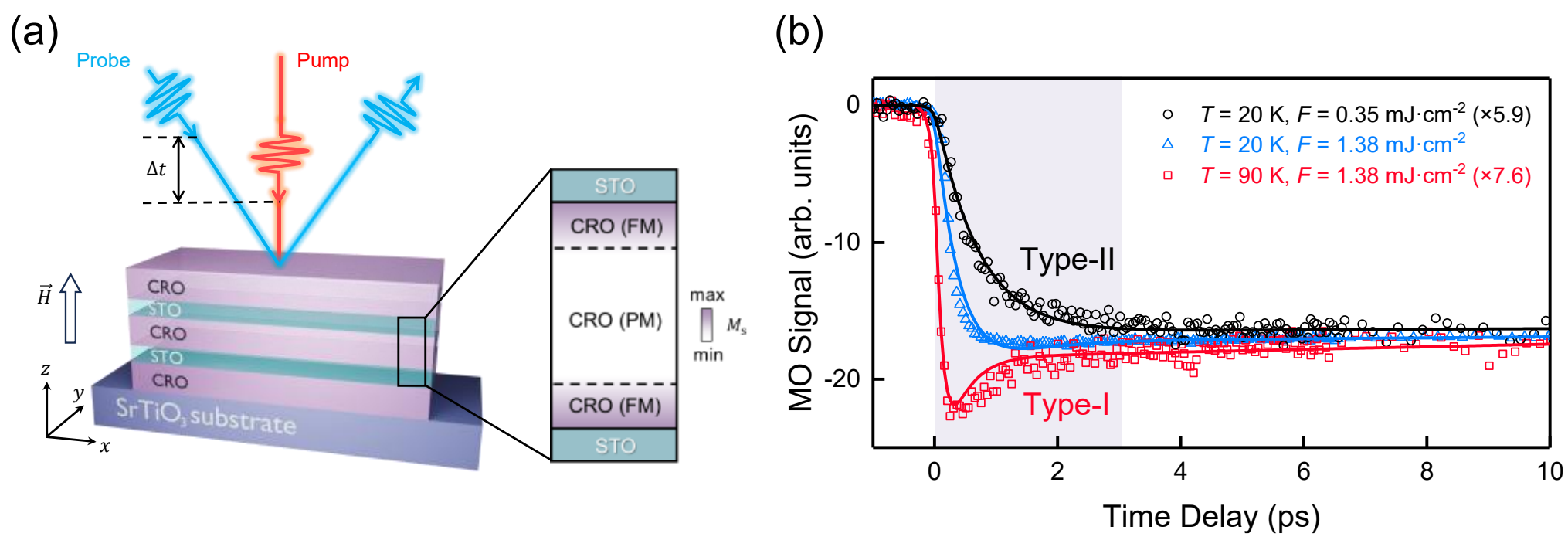

**FIG. 1. Experimental setup and ultrafast demagnetization in CRO/STO SLs. (a)** The left panel is the schematic of the ultrafast all-optical pump-probe setup. The laboratory coordinate system (*xyz*) is defined as shown. Pump pulses at 780 nm and probe pulses at 390 nm propagate along the *z*-axis and incident normally onto the SL surface (shown at a slight angle in the schematic for clarity). A static external magnetic field is applied along the *z*-axis. The right panel is schematic of the gradient-magnetism layered distribution. The gradient magnetism is induced by symmetry mismatch, where ferromagnetic (FM) order is robust at the interfaces and decays towards the paramagnetic (PM) center. The minimum saturation magnetization ($M_S$) is 0 and represents the paramagnetic state. **(b)** Time-resolved magneto-optical (MO) signals for the SL acquired under varying temperatures and pump fluences. Time zero (time delay = 0) corresponds to the onset of the MO response. Here “Type-I” and “Type-II” are used only as phenomenological labels to describe the temporal profiles of the demagnetization curves.

**Figure 2**

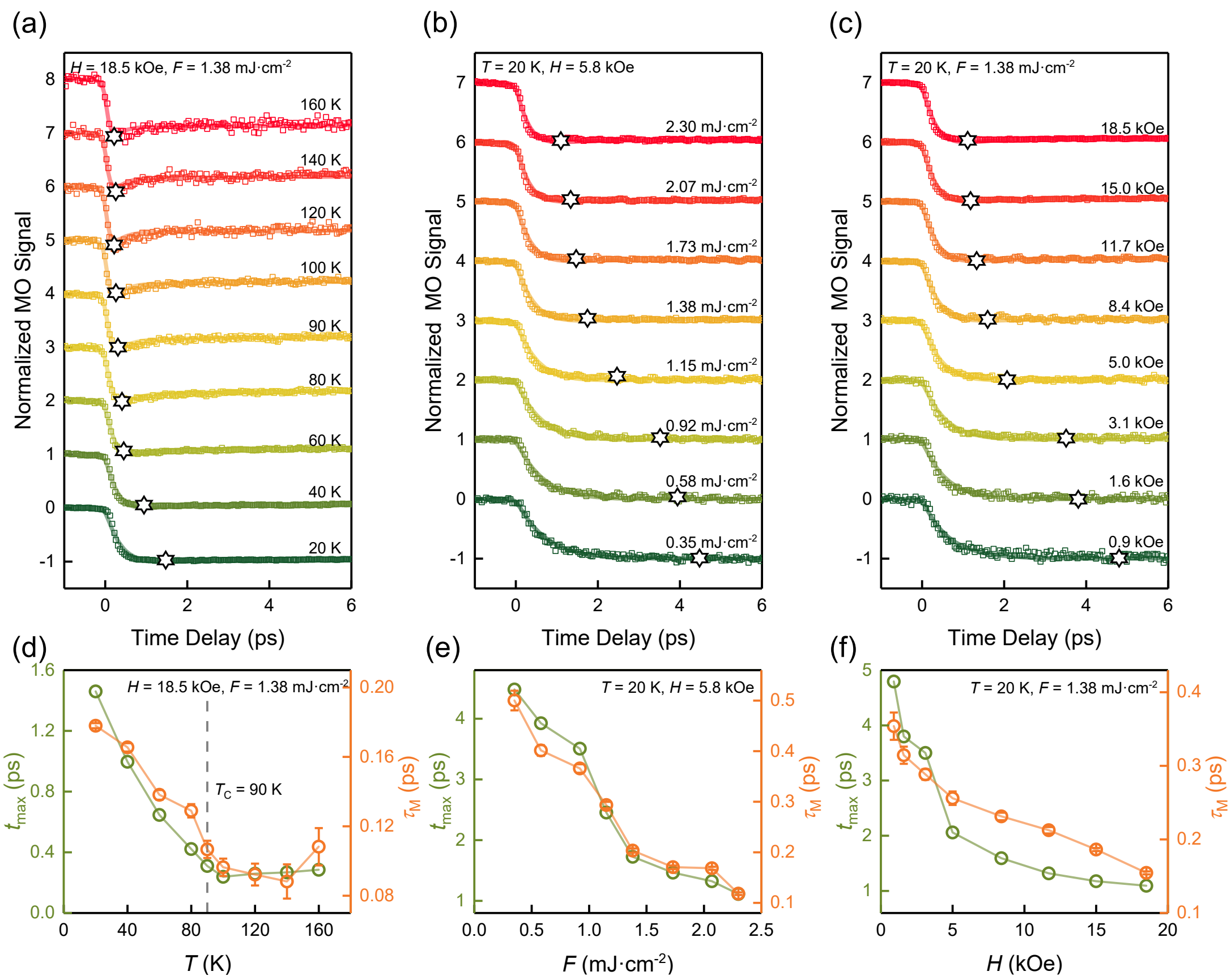

**FIG. 2. Dependence of ultrafast demagnetization of SLs on temperature, pump fluence and magnetic field.** TR-MOKE traces measured under varied **(a)** temperature ($T$), **(b)** pump fluence ($F$), and **(c)** magnetic field ($H$), with remaining parameters held constant in each case. Experimental data (open squares) are fitted (solid lines) using Eq. (1) to extract the spin relaxation rate; the stars indicate the time to maximum demagnetization ($t_{\max}$). Traces are offset vertically for clarity. The dependence of $t_{\max}$ (green) and $\tau_{\mathrm{M}}$ (orange) on each control parameter is summarized in panels **(d)**, **(e)**, and **(f)**, respectively. The signals are min-max normalized to [0, 1] to facilitate intuitive comparison of the characteristic timescales.

**Figure 3**

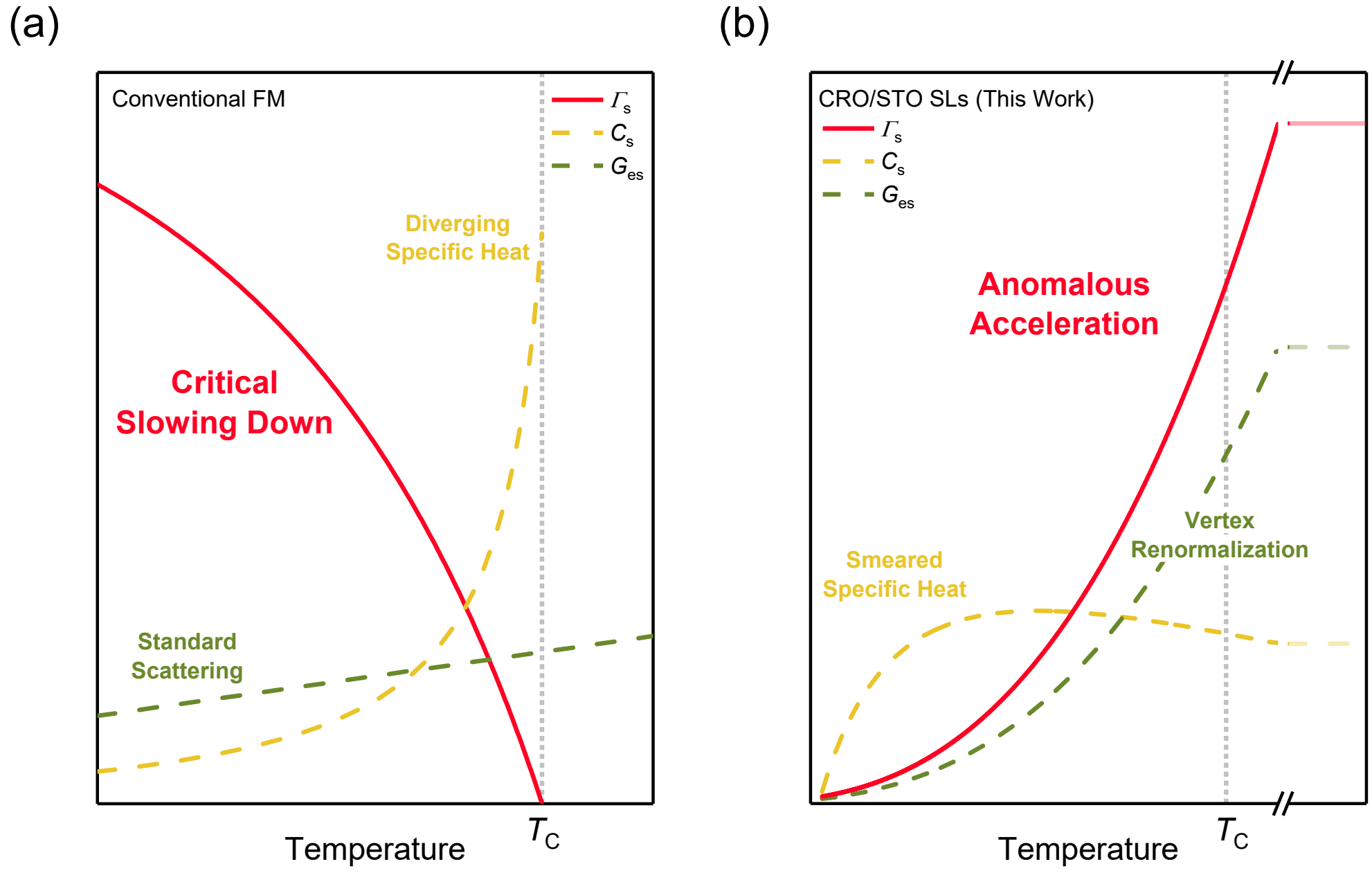

**FIG. 3. Microscopic mechanism of ultrafast demagnetization.** Schematic illustration comparing the governing dynamics in **(a)** a conventional ferromagnet and **(b)** the CRO/STO superlattice (SL). **(a)** In conventional ferromagnets, the divergence of the spin specific heat $C_s$ (yellow dashed line) near the Curie temperature $T_C$ creates a thermodynamic bottleneck. This leads to critical slowing down, causing the spin relaxation rate $\Gamma_s$ (red solid line) to decrease. **(b)** In CRO/STO SLs, gradient magnetism suppresses the divergence of $C_s$ , allowing the spin-fluctuation-driven vertex renormalization ($G_{es}$) to dominate the critical dynamics. Far above $T_C$ (indicated by the axis break), the saturation of $G_{es}$ and $C_s$ imposes a thermalization bottleneck, leading to a constant asymptotic limit of $\Gamma_s$. Note that the curves are schematic representations scaled from the theoretical model to qualitatively illustrate the dynamic trends.

**Figure 4**

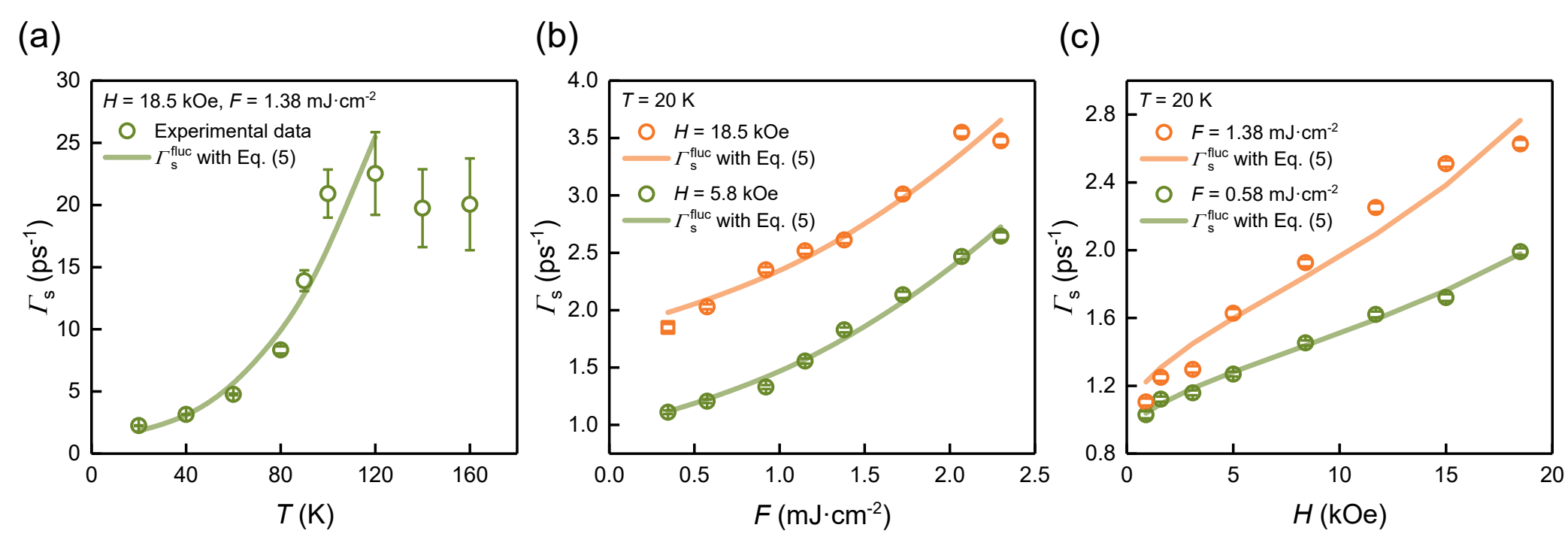

**FIG. 4. Dependence of the spin relaxation rate on external parameters (*T*, *F* and *H*).** The spin relaxation rate $\Gamma_s$ extracted from fits to the TR-MOKE data is plotted as a function of **(a)** temperature ($T$), **(b)** pump fluence ($F$), and **(c)** magnetic field ($H$) (open symbols). The solid-colored lines represent fits using the theoretical model for the fluctuation-mediated relaxation rate, $\Gamma_s^{\mathrm{fluc}}(y, t)$, given by Eq. (5).